\documentclass[journal]{IEEEtran}

\ifCLASSINFOpdf
\else
   \usepackage[dvips]{graphicx}
\fi
\usepackage{url}

\usepackage{graphicx}
\usepackage{epstopdf}
\usepackage[cmex10]{amsmath}
\usepackage{cases}
\usepackage[tight,footnotesize]{subfigure}
\usepackage{amsthm} 
\usepackage{cite}
\usepackage{amssymb}
\usepackage{algorithm}
\usepackage{algorithmic}
\usepackage{multirow}
\usepackage{amsmath}
\usepackage{xcolor}
\usepackage{CJK}
\usepackage{subeqnarray}
\usepackage{cases}
\usepackage{enumerate}

\newtheorem{theorem}{\hskip\parindent\bf{Theorem}}
\newtheorem{lemma}{\hskip\parindent\bf{Lemma}}

\begin{document}
\vspace{-5mm}
\title{On the Generalization and Advancement of Half-Sine-Based Pulse Shaping Filters for Constant Envelope OQPSK Modulation}

\author{Pengcheng Mu,~\IEEEmembership{\normalsize Member,~IEEE,} Yan Liu, Zihao Guo, Xiaoyan Hu$^*$,~\IEEEmembership{\normalsize Member,~IEEE,} and Kai-Kit Wong,~\IEEEmembership{\normalsize Fellow,~IEEE}
\thanks{
This work is supported in part by the Natural Science Foundation of China (NSFC) under Grant 62071370 and Grant 62201449, and in part by the Key R$\&$D Projects of Shaanixi Province under Grant 2021GY-051 and Grant 2023-YBGY-040. \emph{(Corresponding author: Xiaoyan Hu.)}}

\thanks{P. Mu, Y. Liu, Z. Guo, and X. Hu are with the School of Information and Communication Engineering, Xi'an Jiaotong University, Xi'an 710049, China. (email: pcmu@mail.xjtu.edu.cn, 5563GG@stu.xjtu.edu.cn, gzh19970418@stu.xjtu.edu.cn, xiaoyanhu@xjtu.edu.cn)}
\thanks{K.-K. Wong is with the Department of Electronic and Electrical Engineering, University College London, London WC1E 7JE, U.K. (email: kai-kit.wong@ucl.ac.uk)}
}

\markboth{Journal of \LaTeX\ Class Files}
{Shell \MakeLowercase{\textit{et al.}}: Bare Demo of IEEEtran.cls for IEEE Journals}

\maketitle

\vspace{-10mm}
\begin{abstract}
The offset quadrature phase-shift keying (OQPSK) modulation is a key factor for the technique of ZigBee, which has been adopted in IEEE 802.15.4 for wireless communications of Internet of Things (IoT) and Internet of Vehicles (IoV), etc. In this paper, we propose the general conditions of pulse shaping filters (PSFs) with constant envelope (CE) property for OQPSK modulation, which can be easily leveraged to design the PSFs with CE property. 
Based on these conditions, we further design an advanced PSF called $\alpha$-half-sine PSF.
It is verified that the newly designed $\alpha$-half-sine PSF can not only keep the CE property for OQPSK but also achieve better performance than the traditional PSFs in certain scenarios. Moreover, the $\alpha$-half-sine PSF can be simply adjusted to  achieve a flexible performance tradeoff between the transition roll-off speed and out-of-band leakage.
\end{abstract}


\IEEEpeerreviewmaketitle

\vspace{-3mm}
\section{Introduction}
\IEEEPARstart{A}{s} the applications of Internet of Things (IoT) and Internet of Vehicles (IoV) become widespread nowadays, low power consumption design as a urgent requirement for wireless communications attracts more and more attention from both academia and industry~\cite{7123563,2022IoT,TVT2021_YCWang}. 
It it well known that the system power efficiency could be improved not only from the aspect of upgrading the transceiver circuits but also from refining the baseband signal waveform, where the constant envelope (CE) modulation is preferred for the latter.
The CE property is of great importance in guaranteeing the performance of wireless transmissions considering practical hardware limitations \cite{2018Constant}.
The classical CE modulations include the minimum shift keying (MSK)~\cite{2003MSK}, the Gaussian MSK (GMSK)~\cite{1095089}, and some other continuous phase modulations~\cite{2007Continuous}.
All these CE modulations can be written as the sum of offset quadrature phase shift keying (OQPSK) components with half-sine pulse shaping filter (PSF).
OQPSK is an improved QPSK modulation technique which has been widely used in many wireless communication scenario, such as the GNU radio receiver design~\cite{2021OQPSK} and RadCom for IoV~\cite{2022RadCom}, due to its properties of CE and low peak-to-average power ratio (PAPR). It is known that OQPSK modulation plays an important role in the technology of ZigBee and has been standardized as IEEE 802.15.4 for wireless communications of IoT and IoV, etc.~\cite{9144691}. 

Many existing works focus on designing the PSFs of OQPSK modulation, which is essential to improve its time and/or frequency domain properties.
The inter-symbol-interference and jitter-free OQPSK (IJF-OQPSK) in~\cite{1095739} leverage PSFs such as double-symbol interval raised cosine pulse and double-symbol interval modified raised cosine pulse, etc., to ensure that there is no inter-symbol interference and jitter for OQPSK modulation.
The partial response-IJF-OQPSK (PR-IJF-OQPSK) is a refined technique which can further improve the spectral characteristics of IJF-OQPSK signals by using partial response PSFs{\cite{2006Performance}.
It not only retains the fast roll-off characteristics of the IJF-OQPSK signal spectrum, but also has lower out-of-band leakage than IJF-OQPSK.
The cross-correlation phase shift keying (XPSK) is developed on the basis of IJF-OQPSK modulation~\cite{Kato1983XPSK}, and its PSF adopts four transition functions, which make the envelope fluctuation smaller than that of IJF-OQPSK signal.
Although these existing PSFs in~\cite{1095739,2006Performance,Kato1983XPSK} are capable of reducing the spectrum occupancy of OQPSK modulation, the resulted signals usually have increased PAPR, which is unacceptable in practical hardware-constrained wireless communication systems~\cite{2019The}.

In order to further reduce the PAPR of OQPSK signals, half-sine PSF is commonly used due to its intrinsic CE property.
However, the spectrum function of the half-sine PSF decays quite slow and it may induce larger out-of-band leakage under certain bandwidth constraints, which are challenging issues for designing satisfactory PSFs.
Amoroso~\cite{1093294} proposed some general conditions of PSFs with CE property, and gave a PSF  called sinusoidal frequency shift keying (SFSK) satisfying these conditions. 
Although the proposed SFSK PSF has a faster roll-off property while retaining the CE property, its out-of-band leakage is even worse than the half-sine PSF when the bandwidth constraint is stringent.
More importantly, the proposed general conditions of PSFs in\cite{1093294} can only be utilized to verify whether specific PSFs meet the CE property while it is difficult and almost impossible to design PSFs with CE property based on these conditions.

Hence, it is of critical meaning to explore the general conditions of PSFs with CE property which can be leveraged to elaborately design PSFs for CE OQPSK modulation, with the aims of effectively improving the attenuation speed of PSFs' spectrum function and reduce the out-of-band leakage.  These conditions are attractive for the generalization of OQPSK as well as improving the system performance.
Inspired by the above motivations, this paper has the following contributions:
\begin{itemize}
  \item \textbf{Generalization:} The general conditions of PSFs for OQPSK modulation with CE property are concluded which provide meaningful insights for designing the PSFs in wireless communications.
  \item \textbf{Advancement:} An advanced $\alpha$-half-sine PSF is designed based on the proposed general conditions, which is capable to achieve a flexible performance tradeoff between the transition roll-off speed and out-of-band leakage.
  \item \textbf{Verification:} The performance improvement and flexibility of the newly designed $\alpha$-half-sine PSF are verified in comparison with the traditional benchmarks.
\end{itemize}


\vspace{-1mm}
\section{PSFs for OQPSK Modulation}\label{sec:Half}
OQPSK is an improved QPSK modulation technique which only has 0$^\circ$ and 90$^\circ$ phase jumps and overcomes the 180$^\circ$ phase jump of QPSK, and thus has lower PAPR.
The flow chart of OQPSK modulation for wireless communications is shown in Fig.~\ref{fig:flow}, which indicates that the PSFs play an important role in guaranteeing the CE property of OQPSK modulation.
\vspace{-4mm}
\begin{figure}[!htbp]
\centering
\includegraphics[width=3in]{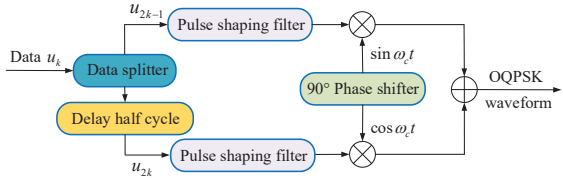}
\caption{The flow chart of OQPSK modulation for wireless communications.}
\label{fig:flow}
\end{figure}

Half-sine PSF is widely leveraged for OQPSK modulation thanks to its CE property, which can facilitate signal processing for wireless communications and acts as an essential part for the technique of ZigBee in standard of IEEE 802.15.4~\cite{9144691}.
The function of half-sine PSF is given as
\begin{equation}
h_1(t)=\left\{\begin{array}{lr}
\cos \left(\frac{\pi t}{2 T}\right), & -T < t < T, \\
0, & \text { otherwise },
\end{array}\right.
\label{con:halfsine}
\end{equation}
with $T$ being the range of the half-sine PSF  and its spectrum function can be written as
\begin{equation}
H_1(\omega) =\int_{-\infty}^{\infty} h_1(t) e^{-j \omega t} d t=\frac{\frac{\pi}{T} \cos \omega T}{\frac{\pi^{2}}{4 T^{2}}-\omega^{2}},
\label{con:half-sine spectrum}
\end{equation}
which is the Fourier transform of $h_1(t)$  with $\omega$ being the angular frequency.
It is easy to observe that the spectrum function $H_1(\omega)$  decays approximately in the form of  $\omega^2$.

Although the half-sine PSF is a common option for OQPSK modulations in wireless communications, it has an inherent drawback with slow attenuation at the transition band.
In~\cite{1093294}, the SFSK PSF is proposed which is expressed as
\begin{equation}
h_2(t)=\left\{\begin{array}{lr}
\cos \left(\frac{\pi t}{2 T}-\frac{1}{4} \sin  \frac{ 2 \pi t}{T} \right), & -T < t < T, \\
0, & \text { otherwise,}
\end{array}\right.
\label{con:SFSK}
\end{equation}
by introducing a sinusoidal frequency shift $\frac{1}{4} \sin  \frac{ 2 \pi t}{T}$ so as to achieve good roll-off characteristics at the transition band while retaining the CE property. Similarly to expression \eqref{con:half-sine spectrum}, we can also get the  spectrum of the SFSK PSF denoted as 
\begin{equation}
\begin{aligned}
H_2(\omega) &=2 \int_{0}^{T} \cos \left(\frac{\pi t}{2 T}-\frac{1}{4} \sin  \frac{ 2 \pi t}{T} \right) \cos \omega t d t
\end{aligned}
\end{equation}
which can be numerically obtained by trapezoidal method \cite{6129703}.
\vspace{-6mm}
\begin{figure}[!htbp]
\centering
\includegraphics[width=3in]{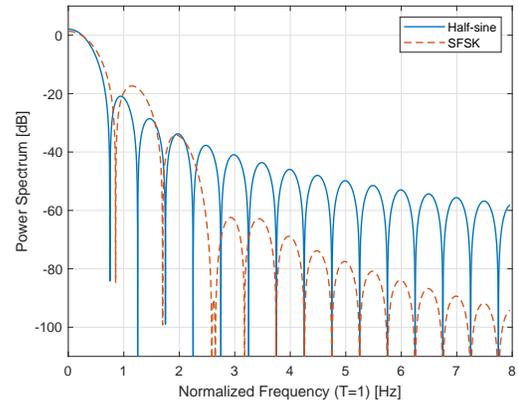}
\caption{The power spectrum of the half-sine and SFSK PSFs.}
\label{fig:SFSK}
\end{figure}

Here, we show the curves of the power spectrum functions for the half-sine and the SFSK PSFs in Fig.~\ref{fig:SFSK}, uniformly expressed as $P_i(\omega)=|H_i(\omega)|^2$ for $i=1,2$, where the frequency $f=\frac{\omega}{2\pi}=\frac{1}{T}$  is normalized with $T=1$ without loss of generality.
It can be seen from Fig.~\ref{fig:SFSK} that the attenuation speed of the SFSK PSF is much faster than that of the half-sine PSF.
However, the bandwidths for the first few sidelobes of the SFSK PSF are much wider than those of the half-sine PSF. Hence, in the scenarios with rigid transition bandwidth requirements, the out-of-band leakage of the SFSK PSF would be much serious than that of half-sine PSF, which is also a quite challenging  issue to be fixed for designing PSFs.

Based on the above overviewed PSFs, we can conclude that the design of PSFs for OQPSK modulation should focus on solving three important challenges:
\begin{enumerate}
  \item Exploring the general conditions of PSFs with CE property for OQPSK modulation, to facilitate the PSF design;
  \item Speeding up the transition decay of PSFs to improve the roll-off property for OQPSK modulation;
  \item Reducing the out-of-band leakage of PSFs to save energy of the main lobe for OQPSK modulation.
\end{enumerate}

\vspace{-2mm}
\section{PSF Design for CE OQPSK Modulation}\label{sec:New}
In this section, we first present the general conditions of  PSFs with CE property for OQPSK modulation based on the traditional half-sine PSF.
Furthermore, a new advanced PSF with more flexible filtering features is  proposed to improve the performance by achieving a desirable tradeoff between the transition roll-off speed and out-of-band leakage.  

\vspace{-3mm}
\subsection{General Conditions of PSFs with CE Property}
As we mentioned before, the general conditions of PSFs with CE property is of great importance to facilitate the design of PSFs for OQPSK modulation.
Although Amoroso proposed a condition of PSFs with CE property given below~\cite{1093294}
\begin{equation}
h^2(t)+h^2(t-T)=1,  \quad 0 < t < T,
\label{con:goal}
\end{equation}
with a  normalized envelope amplitude, it is easy to note that this condition can only be leveraged to verify whether any PSFs satisfy the CE property.
We can prove that the half-sine and the SFSK PSFs given in \eqref{con:halfsine} and \eqref{con:SFSK} both satisfy this CE condition. However, this condition has not provided enough insights on further designing PSFs with CE property.

To conclude the general conditions facilitating the design of PSFs, we consider a uniform expression given below 
\begin{equation}
h(t)=\left\{\begin{array}{lr}
\cos g(t), &-T < t < T, \\
0, &\text{otherwise,}
\end{array}\right.
\label{con:form}
\end{equation}
where $h(t)$  is the half-sine PSF when $g(t)=\frac{\pi t}{2 T}$ and $h(t)$ is the SFSK PSF when $g(t)=\frac{\pi t}{2 T}-\frac{1}{4} \sin  \frac{ 2 \pi t}{T}$.
It is easy to observe that the properties of $g(t)$ greatly affect the performance of $h(t)$ and can provide insights for designing $h(t)$. 

As with the half-sine PSF, $h(t)$ is usually set as a continuous and even function.
Hence, we propose the Theorem \ref{theorem1} given below to show the general conditions for PSF \eqref{con:form} with CE property in the form of an even and continuous function. 

\begin{theorem}\label{theorem1}
For an even-function PSF $h(t)$ in form (\ref{con:form}), if $g(t)$ satisfies
\begin{equation}
\left\{\begin{array}{llr}
g(t) + g(T-t)= \frac{2k+1}{2}\pi, &  k \in \mathbb{Z}, \\
\cos g(T)=0, &
\end{array}\right.
\label{con:condition}
\end{equation}
then $h(t)$ is continuous and the CE property of $h(t)$ can be guaranteed as in formula (\ref{con:goal}).
\end{theorem}

\textit{Proof}:
As $h(t)$ in the form of \eqref{con:form} is an even function, $g(t)$ should be an even or odd function on the interval $(-T,T)$, then we only need to consider $h(t)$ on the interval $(0,T)$ while the other half of the interval can be obtained symmetrically.

Based on the uniform expression of $h(t)$ in (\ref{con:form}), the CE condition in (\ref{con:goal}) can be equally rewritten as
\begin{equation}
\cos g(t)=\pm \sin g(t-T),~0 < t < T,
\end{equation}
which further gives
\begin{equation}\label{gt1}
\cos g(t)= \sin \pm g(T-t)
=\pm \sin  g(T-t),~0 < t < T,
\end{equation}
considering the fact that  $g(t)$ is an even or odd function.
Combining the properties of trigonometric functions, condition \eqref{gt1} can be equally re-expressed as
\begin{equation}
g(t) \pm g(T-t)= \frac{2k+1}{2}\pi, \quad k \in \mathbb{Z}.
\label{con:gen}
\end{equation}
However, if $g(t) - g(T-t)= \frac{2k+1}{2}\pi$ for $k \in \mathbb{Z}$, $g(t)$ is not continuous at $t=\frac{T}{2}$, and thus $h(t)$ becomes discontinuous at $t=\frac{T}{2}$ as well.
In order to maintain the continuity of $h(t)$ at the point of $t=\frac{T}{2}$, we have
\begin{equation}
g(t) + g(T-t)= \frac{2k+1}{2}\pi, \quad k \in \mathbb{Z}.
\end{equation}
In addition, to guarantee the continuity of $h(t)$ at the point of $t=T$, the condition $\cos g(T)=0$ should be hold for sure.

Based on the above derivation, we get the general conditions of  PSFs with CE property as shown in (\ref{con:condition}).
It is clear that if $g(t)$ satisfies the conditions in \eqref{con:condition}, the PSF $h(t)$ in the form of \eqref{con:form} is continuous with CE property. \qed

Without of losing generality, we next consider a simple case with $k=0$ in \eqref{con:condition} to design the PSFs with CE property which should satisfy the following conditions 
\begin{equation}
\left\{\begin{array}{lr}
g(t) + g(T-t)= \frac{\pi}{2}, &\\
\cos g(T)=0.&
\end{array}\right.
\label{con:constant envelope}
\end{equation}

\begin{lemma}\label{lemma1}
For designing a PSF with smooth function $h(t)$, the derivative of $g(t)$ should have the following property
\begin{equation}
\lim _{t \rightarrow T^{-}}  g^{\prime}(t)= 0.
\label{con: derivative continuous}
\end{equation}
\end{lemma}

\textit{Proof}:
To obtain a smooth PSF function $h(t)$,  the derivative of $h(t)$, i.e., $h^{\prime}(t)=-g^{\prime}(t)\sin g(t)$, should be continuous. Hence, $h^{\prime}(t)$ need to be zero at $t=0^+$ and $t=T^-$, which is equivalent to the following conditions
\begin{equation}
\left\{\begin{array}{c}
-\lim\limits_{t \rightarrow 0^{+}} g^{\prime}(t) \sin g(t) =0, \\
-\lim\limits_{t \rightarrow T^{-}} g^{\prime}(t) \sin g(t) =0.
\end{array}\right.
\label{con:derivative}
\end{equation}

According to the first condition in (\ref{con:constant envelope}), we have
\begin{equation}
\left\{\begin{array}{l}
\hspace{-2mm}\lim\limits_{t \rightarrow 0^{+}} \sin g(t)=\lim\limits_{t \rightarrow 0^{+}} \sin\left(\frac{\pi}{2}-g(T-t)\right)=\cos g(T)= 0,\\
\hspace{-2mm}\lim\limits_{t \rightarrow 0^{+}}  g^{\prime}(t)=\lim\limits_{t \rightarrow T^{-}} [\frac{\pi}{2}- g(t)]^{\prime}=-\lim\limits_{t \rightarrow T^{-}}  g^{\prime}(t),
\end{array}\right.
\end{equation}
which means that the first condition of \eqref{con:derivative} is always satisfied. 

However, the second condition in (\ref{con:constant envelope}) leads to
\begin{equation}
\lim\limits_{t \rightarrow T^{-}} \sin g(t)=\lim\limits_{t \rightarrow T^{-}} \pm\sqrt{1-\cos g(t)}=\pm1\neq 0,
\end{equation}
then we should have $\lim\limits_{t \rightarrow T^{-}}  g^{\prime}(t)= 0$ in order to make the second condition in \eqref{con:derivative} satisfied.

Hence, we can conclude that the conditions in (\ref{con:derivative}) that guarantee the smooth property of $h(t)$ can be equivalently simplified as $\lim\limits_{t \rightarrow T^{-}}  g^{\prime}(t)= 0$. \qed

\vspace{-1mm}
\subsection{Advanced $\alpha$-half-sine PSF}
In this part, we design a new type of PSF called $\alpha$-half-sine PSF based on the general conditions proposed in \eqref{con:condition},
which is an advanced half-sine PSF that can not only keep the CE property but also achieve great performance improvement compared with the traditional half-sine PSF.

The $\alpha$-half-sine PSF is defined as 
\begin{equation}
h_{\alpha}(t)=\left\{\begin{array}{lr}
\cos g_{\alpha}(t), &-T < t < T, \\
0, &\text{otherwise},
\end{array}\right.
\end{equation}
with an elaborately designed $g_{\alpha}(t)$ given below
\begin{equation}
g_{\alpha}(t)=\left\{\begin{array}{lr}
\left(\frac{\pi |t|}{{\beta} T}\right)^{\alpha}, & |t| \leq \frac{T}{2}, \\
\frac{\pi}{2}-\left(\frac{\pi(T-|t|)}{\beta T}\right)^{\alpha}, & \frac{T}{2}<|t|<T.
\end{array}\right.
\end{equation}
where $\alpha$ is an adjustable parameter that can be flexibly chosen to achieve a desirable performance tradeoff between the transition roll-off speed and out-of-band leakage.
In order to make sure that $g_{\alpha}(t)$ is continuous at $t = \pm \frac{T}{2}$, we have $\beta=\frac{\pi}{2}\left( \frac{4}{\pi} \right)^{\frac{1}{\alpha}}$.
It is easy to verify that $g_{\alpha}(t)$ satisfies the CE conditions given in (\ref{con:constant envelope}).
In addition, it is obvious that $h_{\alpha}(t)$ becomes the traditional half-sine PSF when $\alpha=1$.

\begin{lemma}
For obtaining a smooth PSF $h_{\alpha}(t)$, the adjustable parameter $\alpha$ should be set as $\alpha> 1$.
\end{lemma}

\textit{Proof}: According to Lemma \ref{lemma1}, the derivative of $g_{\alpha}(t)$ should satisfy that $\lim \limits_{t \rightarrow T^-} g_{\alpha}^{\prime}(t)=0$. As $h_{\alpha}(t)$ is an even function, then we only need to consider the case on the interval $(0,T)$.  $g_{\alpha}(t)$ on  $(0,T)$ is given as
\begin{equation}
g_{\alpha}^{\prime}(t)=\left\{\begin{array}{lr}
\frac{\pi \alpha}{\beta T}\left(\frac{\pi t}{\beta T}\right)^{\alpha-1}, & 0 <t \leq \frac{T}{2}, \\
\frac{\pi \alpha}{\beta T}\left(\frac{\pi(T-t)}{\beta T}\right)^{\alpha-1}, & \frac{T}{2}<t<T.
\end{array}\right.
\end{equation}
Hence, we can obtain the following results that
\begin{equation}
\lim \limits_{t \rightarrow T^-} g_{\alpha}^{\prime}(t)=\left\{\begin{array}{lr}
\infty, & 0<\alpha<1, \\
\frac{\pi \alpha}{\beta T}=\frac{\pi }{2 T}, &\alpha=1, \\
0, &\alpha>1,
\end{array}\right.
\end{equation}
from which we can easily get the meaningful adjustable range $\alpha> 1$ to guarantee the smooth requirement of $h_\alpha(t)$. \qed

The value of $\alpha$ has great effects on the performance of PSFs for OQPSK modulation, which will be verified in the section of numerical results.

\vspace{-1mm}
\section{Numerical Results}\label{sec:Performance}
In this section, we show the performance improvement and flexibility of the newly designed $\alpha$-half-sine PSF with numerical results. We first analyze the performance of the transition roll-off speed for the spectrum function of the $\alpha$-half-sine PSF, and then the performance of the out-of-band leakage under different bandwidth limitations, both compared with the existing half-sine and SFSK PSFs.
The advantages and applicable scenarios are summarized with the consideration of adjusting the parameter $\alpha$ to achieve flexible tradeoff between the two essential performance metrics.

\vspace{-1mm}
\subsection{Performance of Transition roll-off Speed}
The transition roll-off speed is an important performance metric for designing PSFs, which can be reflected by the spectrum functions of PSFs.
The spectrum functions of the half-sine and the SPSK PSFs are given in Section \ref{sec:Half}.
Here, we first calculate the spectrum function of the $\alpha$-half-sine PSF $h_\alpha(t)$, which is given as
 \begin{equation}
\begin{aligned}
H_\alpha(\omega) &=\int_{-T}^{T} \cos g_\alpha(t) e^{- j \omega t} d t \\
&=2 \int_{0}^{\frac{T}{2}} \cos \left(\frac{\pi t}{a T}\right)^{\alpha} \cos \omega t d t \\
&+2 \int_{\frac{T}{2}}^{T} \cos \left[\frac{\pi}{2}-\left(\frac{\pi(T-t)}{a T}\right)^{\alpha}\right] \cos \omega t d t,
\end{aligned}
\end{equation}
and can be numerically obtained by trapezoidal method \cite{6129703}.
\begin{figure}
    \centering
    \includegraphics[width=3in]{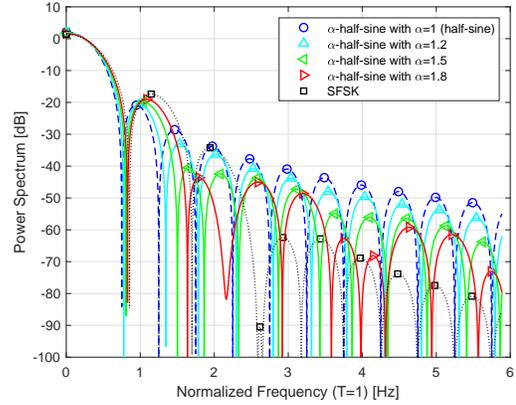}
    \caption{The power spectrums of  different  PSFs.}
    \label{fig:compare}
\end{figure}

The power spectrums of $\alpha$-half-sine PSFs with different values of $\alpha$ in comparison with the traditional half-sine and SFSK PSFs are shown in Fig. \ref{fig:compare}.
From  Fig. \ref{fig:compare}, it is easy to observe that width the main lobe for all the PSFs are almost the same which slightly increases with the increasing of $\alpha$. The half-sine PSF half-sine PSF has the narrowest first sidelobe but with the slowest transition roll-off speed, while the SFSK PSF has the fastest transition roll-off speed but with the widest first sidelobe. In comparison, through adjusting the parameter $\alpha$, the proposed $\alpha$-half-sine PSF can achieve a flexible performance enhancement of the transition roll-off speed.
With the increasing of $\alpha$, the first sidelobe of $\alpha$-half-sine PSF becomes wider, but its the power spectrum decays faster. The $\alpha$-half-sine PSF can achieve better performance of transition roll-off speed than the traditional half-sine PSF by choosing a value of $\alpha>1$.

\subsection{Performance of Out-of-Band Leakage}
The out-of-band leakage is defined as the proportion of out-of-band power, which is also an essential performance indication for designing PSFs.
We can easily calculate the out-of-band leakage of the half-sine, the SFSK, and the proposed $\alpha$-half-sine PSFs under different bandwidth constraints.
A uniform expression of the out-of-band leakage, denoted as $R_o$, under a bandwidth $W$  is given below
\begin{equation}
R_o(\omega)=1-\frac{\int_{-W}^{W} |H(\omega)|^2 d \omega}{\int_{-\infty}^{\infty} |H(\omega)|^2 d \omega},
\end{equation}
which can also represent the decay rate of the power spectrum in some degree and can be numerically calculated by trapezoidal method.
For infinite integrals, we take a quite large bandwidth $W_m$ to approximate the calculation.

\begin{figure}[!htbp]
    \centering
    \includegraphics[width=3in]{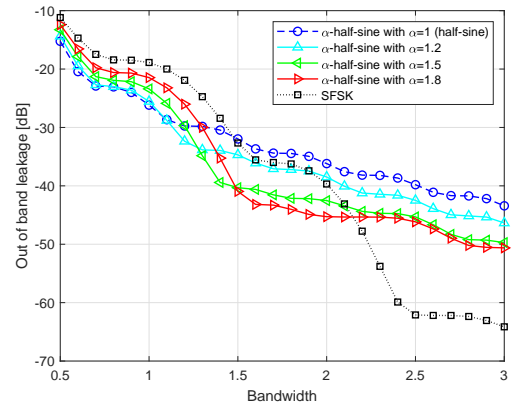}
    \caption{The out-of-band leakage for different PSFs.}
    \label{fig:different pulse shaping filter log}
\end{figure}

The performance of the out-of-band leakage for different PSFs is shown in Fig. \ref{fig:different pulse shaping filter log} versus the normalized bandwidth.
It can be seen that the out-of-band leakage is larger under narrower bandwidth especially with larger value of $\alpha$ for the $\alpha$-half-sine PSF. However, it decays faster, and the out-of-band leakage under wider bandwidths becomes smaller, and the reduction is obvious for larger value of $\alpha$.
We can observe that the SFSK PSF has largest out-of-band leakage for narrow bandwidth and smallest for wide bandwidth.
For the bandwidth constraints between 1.5 and 2, the out-of-band leakage of $\alpha$-half-sine PSF is obviously smaller than half-sine and SFSK PSFs. 
Considering the limited spectrum resources, the proposed $\alpha$-half-sine PSF is of practical meaning in achieving desirable performance under strict bandwidth constraints, through flexibly adjusting the  parameter $\alpha$.

\vspace{-2mm}
\subsection{performance of PAPR}

It is easy to note that the proposed $\alpha$-half-sine PSFs are designed as analog filters, while numerical PSFs are widely used before Digital-to-Analog Convertors (DACs) in practical wireless communication systems.
It is true that the proposed analog PSFs can be simply converted into numerical filters by sampling.
In fact, after the signal passes through the sampled numerical filter, it often needs to go through another filter which is usually an ordinary low-pass filter (LPF). 

Here we consider that the OQPSK signal passes through a finite-length LPF after the sampled digital PSF, and then evaluate the PAPRs of the time-domain signals in Fig.~\ref{fig:papr_time}.
Let's denote the interpolation multiple of the LPF as $N_0=5$, the cut off frequency as $\frac{\pi}{N_0}$, and the sampling points as $-T,$ $\frac{(-N_0+2) T}{N_0}, \cdots, 0, \cdots, \frac{(N_0-2) T}{N_0}$.
The LPF coefficients are set as $h_{-K}, h_{-K+1}, \cdots, h_{0}, 1, \cdots, h_{K}$} with $K=50$, where
\begin{equation}
  h_i=\frac{\pi}{N_0} \frac{\sin \frac{i \pi}{2 N_0}}{\frac{i \pi}{2 N_0}}, \ \forall i=-K, \cdots, 0,  \cdots, K.
\end{equation}
The interpolation multiple of the sampled PSF is set as $N$, and sampling method is the same as that of the LPF.
\vspace{-2mm}
\begin{figure}[!htbp]
    \centering
    \includegraphics[width=3in]{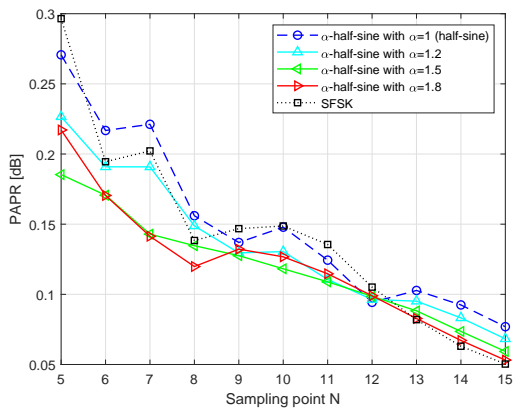}
    \caption{PAPR of OQPSK signal after digital pulse shaping filter and low pass filter for different $N$ ($N_0=5$).}
    \label{fig:papr_time}
\end{figure}

It can be seen from Fig.~\ref{fig:papr_time} that the PAPR performance of the designed $\alpha$-half-sine PSF is better than half-sine PSF and the SFSK especially for low interpolation multiple and with larger $\alpha$. All the PAPR curves show overall downward trends versus interpolation multiple $N$, which indicates that the CE property cannot be held perfectly because of the sampling.

\section{Conclusion}\label{sec:Conclusion}
In this paper, we have presented the general conditions of PSFs with CE property for OQPSK modulation on the basis of the half-sine PSF.
Moreover, an advanced $\alpha$-half-sine PSF keeping the CE property has been elaborately designed based on the proposed general conditions. 
It is verified that the $\alpha$-half-sine PSF decays faster than the traditional half-sine PSF and its out-of-band leakage is much smaller than both the half-sine and SFSK PSFs under certain bandwidth constraints, which can be flexibly adjusted to achieve desirable performance improvement.

The performance of PAPR show that the CE property cannot be guaranteed if the designed analog PSFs are directly converted to numerical filters by sampling.
Hence, it is necessary to optimize the numerical PSFs based on the proposed PSFs by making a compromise between the PAPR performance and bandwidth requirement, which is of practical meaning for utilizing the technology of ZigBee in IoT and IoV, and will be considered as our future works.

\bibliographystyle{IEEEtran}
\bibliography{IEEEabrv,biblio_rectifier}

\end{document}